\documentclass[aps,prb,twocolumn]{revtex4-2}

\usepackage{graphicx}

\usepackage[breaklinks=true,colorlinks=true,urlcolor=blue,
citecolor=blue,linkcolor=blue,bookmarks=true,bookmarksopen=true]{hyperref}
\usepackage{bookmark}

\begin{document}
	
	\title{Doping controlled Fano resonance in bilayer 1T$ ^{\prime} $-ReS$ _{2} $: Raman experiments and first-principles theoretical analysis}

	\author{Subhadip Das$^1$, Suchitra Prasad$^2$, Biswanath Chakraborty$^{1,3}$, Bhakti Jariwala$^4$, Sai Shradha$ ^{4} $, D. V. S. Muthu$^1$, Arnab Bhattacharya$^4$, U. V. Waghmare$^2$ and A. K. Sood$^1$}
	
	
	\email{asood@iisc.ac.in}
	\affiliation{$^1$Department of Physics, Indian Institute of Science, Bangalore 560012, India\linebreak$^2$ Theoretical Sciences Unit, Jawaharlal Nehru Centre for Advanced Scientific Research, Bangalore 560064, India \linebreak $ ^{3} $ Present address: Department of Physics, Indian Institute of Technology Jammu, Jammu-181221,J\&K, India\linebreak $^4$Department of Condensed Matter Physics and Materials Science, Tata Institute of Fundamental Research, Mumbai 400005, India.}
	
	\begin{abstract}
		
		In the bilayer ReS$ _{2} $ channel of a field-effect transistor (FET), we demonstrate using Raman spectroscopy that electron doping  (n) results in softening of frequency and broadening of linewidth of the in-plane vibrational modes, leaving out-of-plane vibrational modes unaffected. Largest change is observed for the in-plane Raman mode at $\sim$ 151 cm$^{-1} $, which also shows doping induced Fano resonance with the Fano parameter $ 1/q = -0.17 $ at doping concentration of $\sim 3.7\times10^{13}$ cm$^{-2} $. A quantitative understanding of our results is provided by first-principles density functional theory (DFT), showing that the electron-phonon coupling (EPC) of  in-plane modes is stronger  than that of  out-of-plane modes, and its variation with doping is independent of the layer stacking. The origin of large EPC is traced to 1T to 1T$ ^{\prime} $ structural phase transition of ReS$ _{2} $ involving in-plane displacement of atoms whose instability is driven by the nested Fermi surface of the 1T structure.  Results are also compared with the isostructural trilayer ReSe$ _{2} $.

	\end{abstract}
	
	\maketitle
	\section{Introduction}
	ReS{$_2$}  is a member of group-VII transition metal dichalcogenide (TMD) family with distorted octahedral (1T$ ^{\prime} $) crystal structure \cite{lamfers1996crystal,wildervanck1971dichalcogenides}. Due to Peierls distortion, Re-Re  metal bonds form a chain along the crystallographic \textit{b}-axis \cite{lamfers1996crystal,wildervanck1971dichalcogenides}, which reduces the symmetry of the lattice. ReS$ _{2} $ thus  possesses unique anisotropic electrical and optical properties along its two principle axes \cite{friemelt1993optical,ho2001plane,lin2015single,chenet2015plane,aslan2015linearly}, making it an ideal candidate for strain sensors \cite{yu2016strain}, polarization sensitive photodetectors \cite{zhu2019broadband} and 2D logic circuits \cite{liu2015integrated,kwon2019all}. Due  to very weak interlayer electronic and vibrational coupling, the Raman spectrum of ReS$ _{2} $ does not show thickness dependence \cite{Tongay}. It may be interesting to note some similarity  to the ZnO multi-microspheres mesocrystal, where the mechanical vibrational modes are independent of the size of the individual microspheres, due to propagation of vibrations along the nano-cantilever contacts between the neighboring microspheres \cite{Wu2011}. In addition, the direct bandgap from bulk to monolayer \cite{aslan2015linearly,Tongay} allows few-layer ReS$ _{2} $ to show record high photoresponsivity compared to other 2D materials \cite{liu2016high2}. With excellent transistor performance \cite{corbet2014field,shim2016high,zhang2015res2} and doping-induced transition to a metallic state at $\sim$ 2 K \cite{doi:10.1021/acs.nanolett.5b04100}, studies of electron-phonon coupling (EPC) at high gate bias will provide crucial information about its high field electronic transport properties.      
	
	Raman spectroscopy, a proven non-invasive characterization tool, has been extensively used to study interlayer electronic and vibrational coupling \cite{Tongay}, stacking order \cite{he2016coupling} and crystal vertical orientation \cite{hart2016rhenium} in ReS$ _{2} $. In recent years, it has been used to study EPC by measuring the changes in Raman frequency ($\omega$) and linewidth ($\gamma$) as a function of doping \cite{das2008monitoring,mosbiswa,chakraborty2016electron}. It has been shown that the effect of doping is different for the G and 2D modes of graphene \cite{das2008monitoring}. The symmetry of phonons and electronic states at the conduction band minimum (CBM) and valence band maximum (VBM) play a crucial role in doping induced renormalization of phonons, as demonstrated for monolayer MoS$ _{2} $ \cite{mosbiswa} and black phosphorus \cite{chakraborty2016electron}.

	Fano resonance in Raman spectroscopy emerges from quantum interference between the transition amplitudes of phonon localized states and electronic continuum  \cite{PhysRevB.63.155414,PhysRev.124.1866}. The resulting asymmetric intensity lineshape is defined by the Fano parameter, $ 1/q $, which characterizes coupling strength between the phonon and continuum and hence is a measure of the EPC  \cite{RevModPhys.82.2257}. Semimetals such as bilayer Graphene and TaAs show tunable Fano resonance with $|1/q|\geq$1 \cite{tang2010tunable,xu2017temperature}. For these materials, phonon induces interband electronic transitions, resulting in strong EPC. Similar effect  has also been reported for topological insulators and high-T$ _{c}$ superconductors with relatively small Fano parameter ($\leq$ 0.24) \cite{PhysRevB.81.125120,PhysRevB.91.235438,PhysRevB.91.104510,PhysRevB.93.125135}. Due to considerable  bandgap with low EPC, this effect is not seen in semiconducting TMDs, where tunable Fano resonance can open up a new avenue in the field of sensors, switches, optoelectronics and other novel device applications \cite{luk2010fano,RevModPhys.82.2257}.

	Here we report that with electron doping in bilayer ReS$ _{2} $, the modes with  in-plane vibration show a decrease in phonon  frequency and broadening of linewidth whereas the out-of-plane modes remain unchanged. The in-plane mode at $\sim$151cm$ ^{-1} $ shows carrier-tunable Fano resonance with the largest change in phonon frequency and linewidth compared to other modes. First-principles density functional theory calculations show that the  EPC is higher for the in-plane modes as compared to the out-of-plane modes.  We have determined the stable stacking configuration for bilayer ReS$ _{2} $ and show that the  EPC is similar for all these configurations. To explain our EPC trend, we show that the lattice displacements, which are associated with the transition from 1T to 1T$ ^{\prime} $ phase of ReS$ _{2} $, are among the modes that couple strongly with electrons.

	\section{Methods}
	\subsection{Experimental details}
	ReS$_2$ single crystals \cite{jariwala2016synthesis} were mechanically exfoliated on a 300nm thick oxidized surface of p$ ^{++} $ doped silicon substrate (M/s Nova Electronic Materials). Atomic force microscope (AFM) was employed in tapping mode to determine nanocrystal thickness. Source, drain and gate contacts were drawn out by electron beam lithography followed  by thermal vapor deposition of 5nm and 50nm thick Cr and Au, respectively.  A drop of ionic liquid (IL) (EMIM-TFSI from M/s Sigma-Aldrich) was drop casted  on top of the two-probe device for top gating. A Keithley-2400 source meter was used to perform electrical measurements.  Confocal Raman measurements in back-scattering geometry were performed at room temperature in a commercial Horiba Labram HR-800  spectrometer consisting of a 1800 lines/mm grating coupled with a Peltier cooled CCD as detector.  Linearly polarized lasers of 532 and 660 nm wavelength were focused on the nanocrystal using a 50$\times$ objective of 0.45 NA at incident power less than 0.5 mW. The diameter of the laser spot was $\sim$2 $\mu$m. The resolutions of the captured Raman spectra were  0.52 and 0.35cm$ ^{-1} $ for 532 and 660 nm of laser excitations, respectively. The plane of polarization of the incident laser beam was changed by placing a half-wave plate at the incident path.
	\subsection{First-principles calculations}
	Our first-principles calculations are based on density functional theory (DFT) as implemented in Quantum ESPRESSO package \cite{Giannozzi}. The exchange-correlation energy functional was treated using a Local Density Approximations (LDA) with Perdew-Zunger functional \cite{Zunger.1981}. Ultrasoft pseudopotentials (USPPs) \cite{Vanderbilt.1990} have been used to model the interaction between valence electrons and ionic cores. We have used fully relativistic ultrasoft pseudopotentials to take into  account the spin-orbit coupling (SOC). We used a kinetic energy cut-off of 40 Ry to truncate the plane-wave basis used to represent Kohn-Sham wavefunctions. Structures were relaxed to minimum energy until the Hellman-Feynman forces are less than 15 meV/\AA. Reducing it to 10 meV/\AA \space only marginally changes atomic positions and would not affect any of our results. In the simulation of a bilayer, we have used a vacuum  layer of 10 \AA \space thickness along \textit{z}-direction of the periodic cell to maintain weak interaction between periodic images. We have used fully relativistic ultrasoft pseudopotentials to take into account the spin-orbit coupling (SOC). For the truncation of plane wave basis set, we have used the kinetic energy cut off of 65 Ry. We used uniform meshes of 6$\times$6$\times$6 and 6$\times$6$\times$1 k-points for sample integrations over Brillouin Zones (BZ) of bulk and bilayer ReS$_2$ respectively. We have tested the convergence of our calculations that include spin-orbit interaction, and find that refining the sampling of Brillouin zone from 6$\times$6$\times$1 to 12$\times$12$\times$1 k-point mesh does not change the total energies, and more importantly, the band gap remains the same. Zone-center phonon frequencies were obtained using  DFT linear response as implemented in the Quantum ESPRESSO package. We have used the same numerical parameters and carried out calculations for bulk ReS$_2$ within a generalized gradient approximation (GGA)  of the exchange correlation energy with  PW91 parametrized form \cite{Burke.1998}. The zone-center phonons of bulk ReS$_2$ obtained  within LDA and GGA flavors of DFT  calculations show a good agreement between results of LDA-USPP calculations with experimental values as well as other theoretical calculations reported earlier \cite{feng2015raman}. Secondly, the inclusion of vdW interaction does not affect our estimates of vibrational frequencies notably \cite{feng2015raman}. In the rest of the work presented here, we have thus used LDA-USPP based DFT calculations.

	\section{Results and discussion}
	\subsection{Experimental measurement}
	
	\begin{figure*}[t!]
		\includegraphics[width=1\textwidth]{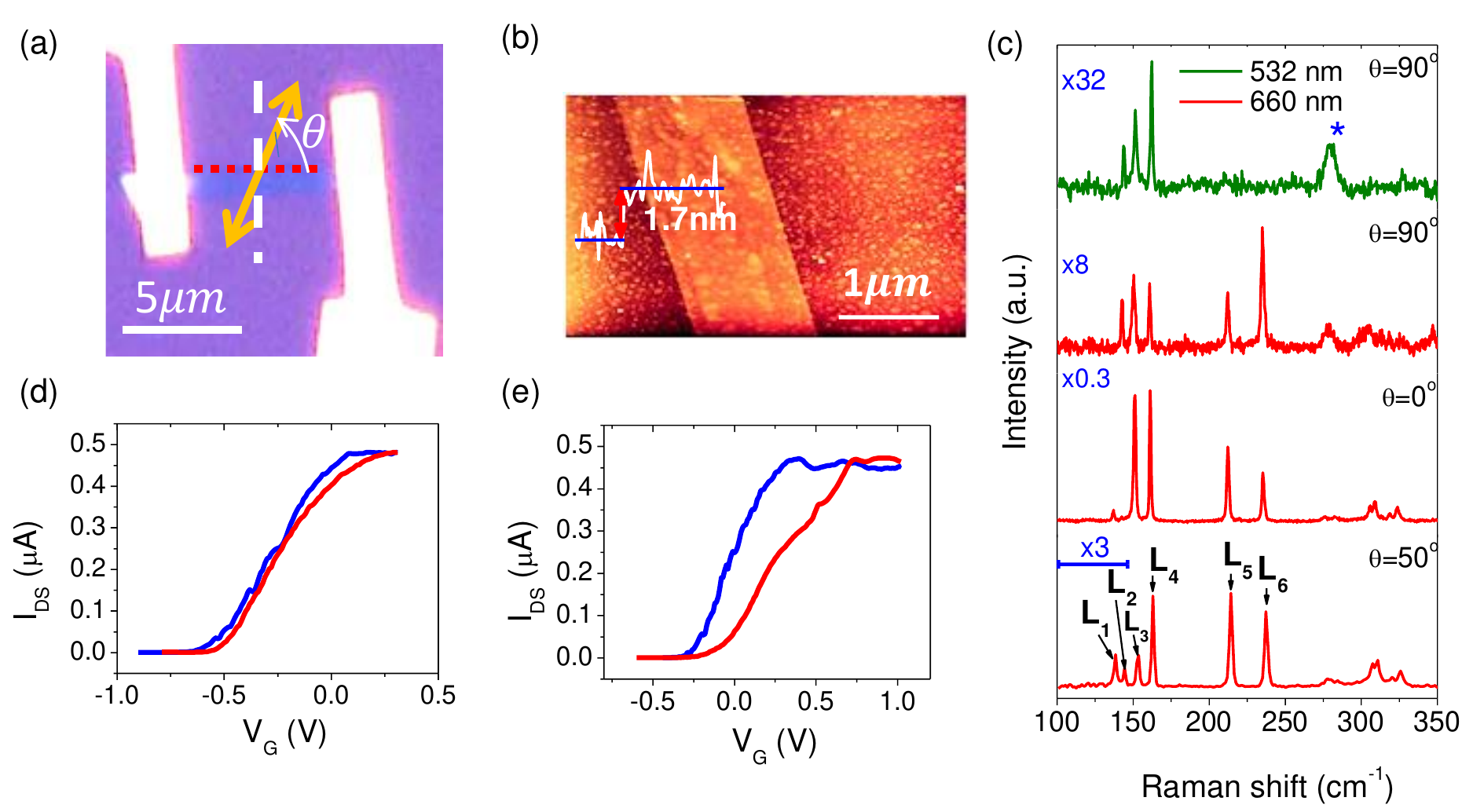}
		\caption{(a) Optical image of the device. The red and white dashed lines indicate the cleaved edge  of the nanocrystal and its normal, respectively. Plane of polarization of the incident laser is represented by the yellow double arrow. (b) AFM height profile of the nanocrystal. (c) Raman spectra of the nanocrystal with 532 and 660 nm of laser excitations at one ($\theta=90^{o}$) and three ($\theta=0^{o}, 50^{o}$ and $90^{o}$) incident polarization angles, respectively with unpolarized collection configuration. Some region of the spectrum at $\theta=50^{o}$  are enhanced (shown by the blue line). The modes are labeled from L$ _{1} $ to L$ _{6} $. For ease of intensity comparison, Raman spectra at $\theta=0^{o}$ and $ 90^{o} $ are re-scaled. The star symbol represents Raman peak from the ionic liquid near 281 cm$ ^{-1} $. (d and e) Drain current (I$ _{DS} $) as a function of gate voltage (V$ _{G} $) at 0.1 V of drain bias. Forward and backward gate voltage sweeps are indicated by the blue and red lines respectively.}
		\label{f1}
	\end{figure*}

	Fig. \ref{f1}(a)  shows the optical micrograph of the two-probe device. The intensity of the Raman modes was changed by rotating the incident laser polarization ($\theta$) with respect to  the cleaved edge of the nanocrystal. To maximize intensity, the scattered spectra were kept unpolarized. AFM height profile measurement confirms $\sim$ 1.7 nm of nanocrystal thickness, which correspond to two layers (Fig. \ref{f1}(b)).  Fig. \ref{f1}(c) shows the ambient Raman spectra of the bilayer ReS$ _{2} $ in the range from 100 to 350 cm$ ^{-1} $ using 532 and 660 nm of laser excitations at one ($\theta=90^{o}$) and three ($\theta=0^{o}, 50^{o}$ and $90^{o}$) polarization angles, respectively. At $\theta=50 ^{o} $, six Raman modes between 130 and 265 cm$ ^{-1} $ are labeled from L$ _{1} $ to L$ _{6} $.  There are four formula units in a unit cell that results in  18 zone-center  Raman modes of A$ _{g} $ symmetry \cite{feng2015raman}. We have followed the convention of  Feng \textit{et al.}  where the modes with dominant out-of-plane and in-plane vibrational weight are called A$ _{g} $-like ( L$ _{1} $ and L$ _{2} $) and E$ _{g} $-like (L$ _{3}$ to L$ _{6} $) respectively \cite{feng2015raman}. Other modes are coupled C$ _{p} $ modes with mixed in-plane and out-of-plane character \cite{feng2015raman}. As a function of $\theta$, the intensity of the out-of-plane modes, L$ _{1} $ and L$ _{2} $ shows two-lobed shape with maxima separated by $\sim80^{o}$ (see section-I of the supplemental material (SM)), in agreement with previous reports \cite{chenet2015plane, doi:10.1021/acsnano.7b05321}. Thus to capture both of these modes with electron doping, Raman spectra is taken at both 0$ ^{o} $ and 90$ ^{o} $ of laser polarization configurations (Fig. \ref{f1}(c)).
	
	Compared to the excitation at 660 nm, the Raman mode intensities are significantly weaker at 532 nm with very weak signal response from L$ _{5} $ and L$ _{6} $ modes at $\theta=90^{o}$ (Fig. \ref{f1}(c)). As previously reported by McCreary et al. \cite{intricate}, Raman intensities depend on the nature of the phonon modes and are not the same at a given  laser wavelength. The Raman modes involving mostly Re dominated in-plane vibrations (L$ _{3} $ to L$ _{6} $) are stronger at 633 nm excitation compared to 515 nm and 488 nm \cite{intricate}. We have used this knowledge to do doping dependence measurements using 660 nm laser excitation.
	
	\begin{figure*}[t!]
		\includegraphics[width=0.9\textwidth]{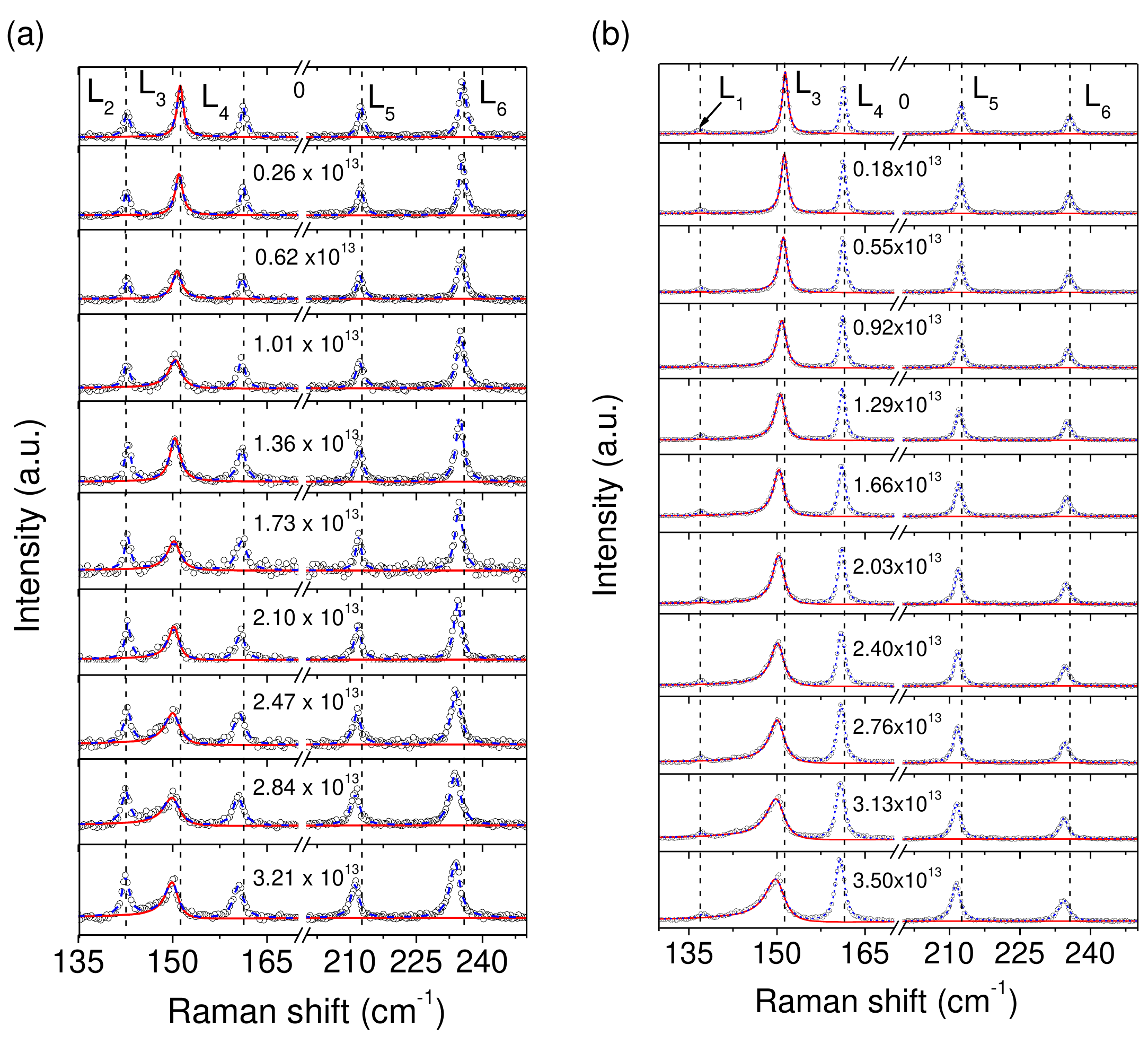}
		\caption{Raman modes at different electron doping concentrations during (a) doping cycle at $\theta=90^{o}$ and  (b) de-doping cycle at $\theta=0^{o}$ . Doping concentrations in cm$ ^{-2} $ units are indicated in the figures. Black circles and blue dashed lines are the experimental data points and their cumulative peak fits respectively. Red solid lines show the BWF profile of the  L$ _{3} $ mode. Black dashed lines are guide to the eye for the Raman peak positions.}  
		\label{f12}
	\end{figure*} 
	
	Consistent with previous reports \cite{jariwala2016synthesis,liu2016high2,lin2015single,corbet2014field,ovchinnikov2016disorder,xu2015sulfur}, the drain current (I$ _{DS} $) as a function of gate voltage (V$ _{G} $)  of the two-probe device (Figs. \ref{f1}(d) and (e))  shows n-type semiconducting behavior. For extended operation in ambient condition, EMIM-TFSI ionic liquid is known to absorb water molecules from air \cite{doi:10.1021/jp3024233} which reduces its electrochemical window \cite{doi:10.1021/je800678e}. Thus to capture the Raman spectra during electron doping and dedoping cycle and avoid any device degradation, we have separately taken two instances (Figs. \ref{f1}(d) and (e)) of the forward and backward voltage sweeps, between which the device was vacuum dried to restore the electrochemical window of IL. We note that the current on-off ratio remains the same ($\sim$ 10$ ^{3} $) for all of the voltage sweeps, implying absence of any device degradation. Although the experimentally measured field-effect mobility values are consistent with previous reports \cite{jariwala2016synthesis,liu2016high2,lin2015single,corbet2014field,ovchinnikov2016disorder,xu2015sulfur}, it decreases from $\sim$ 7.8 cm$ ^{2}/ $V.s to $\sim$ 5.3  cm$ ^{2}/ $V.s with increase in the current threshold voltage (V$ _{Th} $) by $\sim$ 0.4V during doping and dedoping runs in Figs. \ref{f1}(d) and (e), respectively. This effect and the observed hysteresis during doping and dedoping cycle in our measurements can be attributed to the increased electron scattering by the electrostatic disorders induced by IL \cite{Gallagher2015,PhysRevLett.105.036802, ovchinnikov2016disorder}.  Moreover, these disorders also particularly suppress the conductivity of monolayer ReS$ _{2} $ at high gate bias \cite{ovchinnikov2016disorder}. Since the disorders are only limited to gate-solid interface, we have carried out our measurements on a bilayer device, where this effect only saturates the conductivity at high gate bias \cite{ovchinnikov2016disorder} ($ \sim$ 0.1V (0.3V) and 0.33V (0.72V) during forward (return) gate voltage sweeps in Figs. \ref{f1}(d) and (e), respectively).  We note that this saturation of conductivity is due to the distinct band structure of 1T$ ^{'} $-ReS$ _{2} $ with low electron localization length at high gate bias and does not represent any electronic or structural transition from insulating to metallic phase  \cite{ovchinnikov2016disorder}. Throughout the doping range, ReS$ _{2} $ always remains in the semiconducting 1T$ ^{'} $ phase \cite{ovchinnikov2016disorder}. 
	
	\begin{figure*}[t!]
		\includegraphics[width=0.8\linewidth]{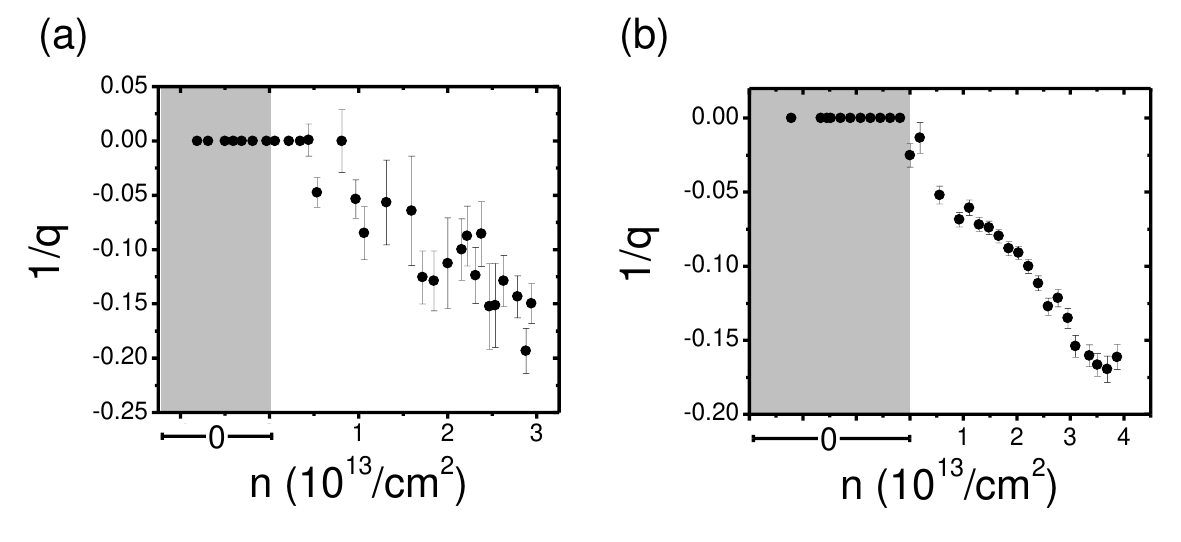}
		\caption{Fano parameter $ (1/q) $ of L$ _{3} $ mode with electron doping at (a) 90$ ^{o} $ and (b) 0$ ^{o} $ of laser polarization during doping and dedoping cycle, respectively. Gray regions indicate zero doped state.}
		\label{f13}
	\end{figure*} 
	
	EMIM-TFSI IL has a reported gate capacitance (C$ _{G} $) of $\sim$ 5.9 $\mu$F/cm$ ^{2}$ at the low frequency regime ($\sim$ 0.02 Hz) \cite{ionic_liq}. To extract the doping concentration (n) from V$ _{G} $, we have used the parallel plate capacitor equation $ n=C_G(V_{G}-V_{Th}) $. Fig. \ref{f12}(a) and (b) compares the Raman spectra in the range from 130 to 265 cm$ ^{-1} $ at $\theta=90^{o}$ and $0^{o}$  at different doping concentrations during doping and dedoping gate voltage sweeps from Fig. \ref{f1}(d) and (e), respectively. At zero doping, all the modes show symmetric Lorentzian lineshape. However with electron doping, the L$ _{3} $ mode becomes increasingly asymmetric and shows broadening on the low frequency side, resembling a Fano lineshape with negative Fano parameter. This effect arises from coupling between the optical phonon mode L$ _{3} $  with electrons in conduction band continuum,
	as the Fermi level  crosses CBM \cite{JOUANNE19751047,PhysRevB.17.1623}. Raman spectral lineshape from Fano resonance is expressed by the Breit-Wigner-Fano (BWF) function,
	\begin{equation}
		I(\omega)=I _{0}\frac{[1+(\omega-\omega_{0})/q\gamma]^{2}}{1+[(\omega-\omega_{0})/\gamma]^{2}},  
		\label{e1}
	\end{equation}
	
	where I$ _{0} $, $\omega_0$, $\gamma$ and $ 1/q $ are, in respective order, the  intensity, bare frequency, linewidth and Fano  parameter, characterizing coupling strength between the phonon and continuum spectra \cite{PhysRevB.8.4734,RevModPhys.46.83}. Thus at each gate voltage we  have fitted the Raman spectra with the BWF function (for L$ _{3} $ mode) together with Lorentzian functions for rest of the four modes in Fig. \ref{f12}, to extract their lineshape  parameters. Figs. \ref{f13}(a) and (b) show two very similar plots of the Fano parameter $ (1/q) $ of the L$ _{3} $ mode with n at $\theta=90^{o}$ and $0^{o}$, respectively. The frequency of the L$ _{2} $ and L$ _{3} $ modes are 142 and 151 cm$ ^{-1} $, respectively.  Thus the asymmetric broadening of the L$ _{3} $ mode on the low frequency side can be analysed more accurately at $\theta=0^{o}$, where L$ _{2} $ has the lowest intensity (Fig. \ref{f1}(c)).   At  n=0 (\textit{i.e.} V$ _{G}\leq$V$ _{Th} $), $1/q \sim$ 0, indicating a symmetric Lorentzian lineshape like all the other modes. With electron doping, the Fano parameter $ |1/q| $ increases almost linearly up to  $\sim$ 0.17 at $n\sim 3.7\times10^{13}$ cm$^{-2} $ (Fig. \ref{f13}(b)). As the Fermi energy increases above the CBM, a higher concentration of electrons causes stronger EPC and thus increasing the Fano parameter.

\begin{figure*}[t!]	
		\centering
		\includegraphics[width=1\textwidth]{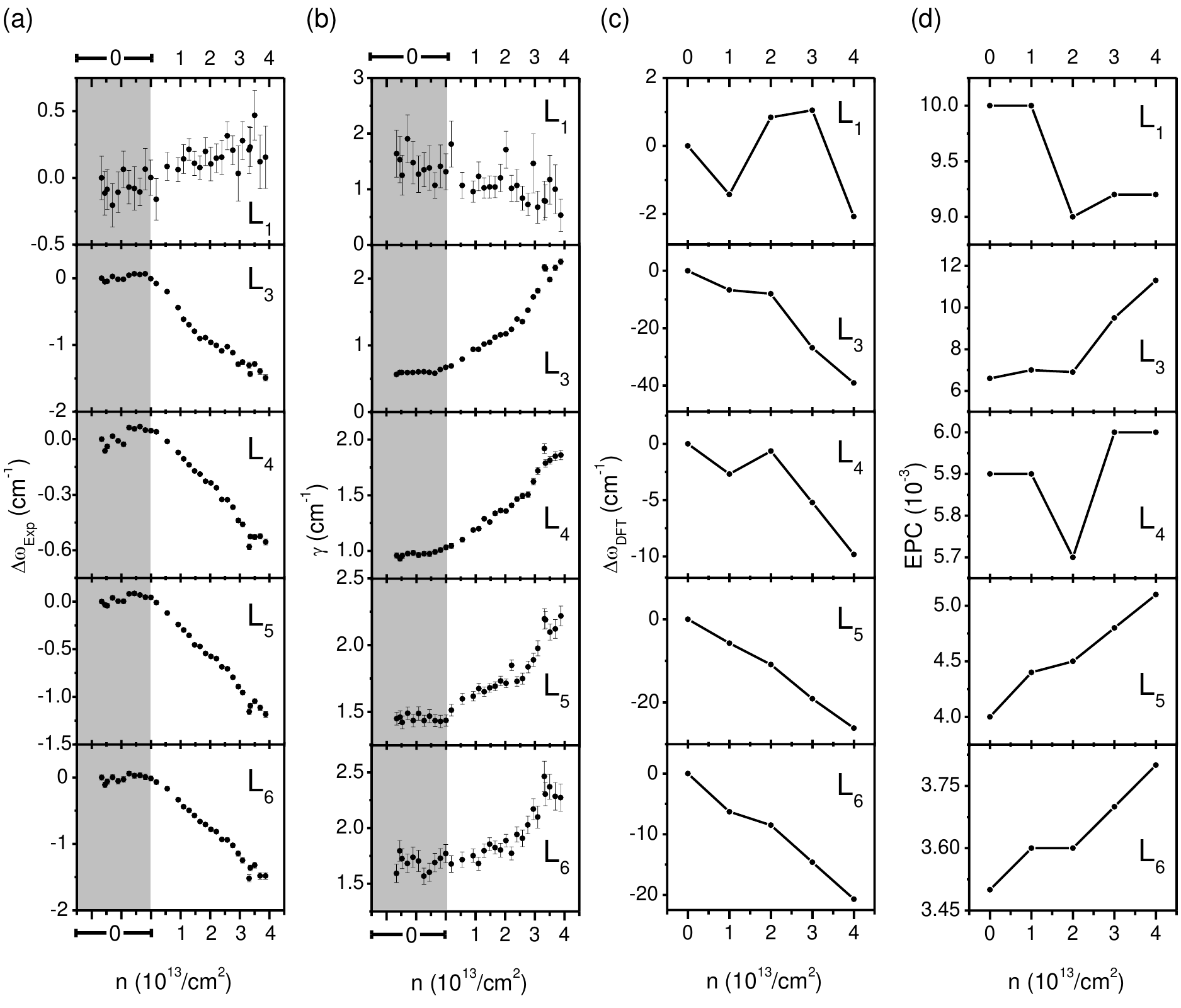}
		\caption{ Experimentally measured  (a) $ \Delta\omega $ ($\omega_{n\neq0}-\omega_{n=0})$ and (b) $\gamma$ with n at $\theta=0^{o}$. Gray regions represent zero doped state. DFT calculated values of (c) $ \Delta\omega $ and (d) EPC as a function of n.}
		\label{f2}
	\end{figure*}

	In section -II and III of the SM (for $\theta=90 ^{o} $) and Figs. \ref{f2}(a) and (b) (for $\theta=0 ^{o} $), the change in frequency from zero doped state, $\Delta\omega (=\omega(n)-\omega(0)$) and $\gamma$ with n are shown. With doping, the E$ _{g} $-like modes (L$ _{3} $, L$ _{4}$, L$ _{5} $ and L$ _{6}$) show phonon softening and linewidth broadening. In contrast, the A$ _{g} $-like modes (L$ _{1} $ and L$ _{2} $) show negligible change.  We note that the doping dependence of the A$ _{g} $ and E$ _{g} $-like modes remains similar in the two perpendicular polarization configurations with the largest change in $\Delta\omega$ and $\gamma$ observed for the L$ _{3} $ mode. Monolayer ReS$ _{2} $ has three excitons at 1.61, 1.68 and 1.88 eV \cite{aslan2015linearly}. For the bilayer, the third exciton energy is red shifted from 1.88 eV to  1.7 eV \cite{aslan2015linearly}. The third highest excitonic level is the one which is relevant for understanding resonance effects using 660 nm (1.88 eV) excitation. Hence, the resonance condition arising from the exciton-3  will not be affected by the doping or polarization of the incident laser. Thus, the doping dependence of the modes  is independent of the laser wavelength or polarization angle. It is interesting to compare these results with phonon renormalization in monolayer MoS$ _{2} $  where the  A$ _{1g} $ mode shows phonon softening and linewidth broadening with negligible effect on the E$ _{2g} $ mode \cite{mosbiswa}. 
	
	In trilayer ReSe$ _{2} $, an isostructural sister compound of ReS$ _{2} $ \cite{lamfers1996crystal,wildervanck1971dichalcogenides}, small change of $\Delta\omega$ and $\gamma$ with electron doping up to 5 $\times$ 10$ ^{13} $ cm$ ^{-2} $ is measured for all of its 14 Raman modes. From DFT calculations we support our experimental results and show that the indirect bandgap  nature of ReSe$ _{2} $  results in weak coupling of electrons with all the phonon modes (see section-IV of the SM).
	
	Non-uniform doping of the devices can also lead to inhomogeneous broadening. This broadening is expected to be similar for all the in-plane Raman modes in ReS$ _{2} $ as well as in ReSe$ _{2} $ whose frequencies are renormalized by the doping. We did not see the asymmetric broadening for other modes. Furthermore, the channel length of our device is $\sim$  6.3 $\mu$m and laser spot diameter is typically of $\sim$ 2 $\mu$m. The doping of the ReS$ _{2} $ channel is uniform as seen by the similar recorded Raman spectra during doping and dedoping cycle at two different spots as shown in Fig. S9 from section-V of the SM. This assures us that the asymmetry observed in the L$ _{3} $ mode on doping is not due to strain or doping inhomogeneity but is intrinsic to ReS$ _{2} $. Notably, non-corroding metal such as Au cannot reach thermodynamic equilibrium through chemical reactions \cite{inzelt2013pseudo}, which can lead to unstable electrode potential in a transistor.  In addition, due to the high gate capacitance of the ionic liquid, the voltage-drop across the drain and source electrodes from drain bias can offset the applied gate voltage across the device channel. Therefore, the actual carrier concentration of the device can differ from the estimated value, especially at a low gate bias.

	\subsection{Theoretical calculations}
	\begin{figure*}[t!]	 
		\includegraphics[width=0.6\textwidth]{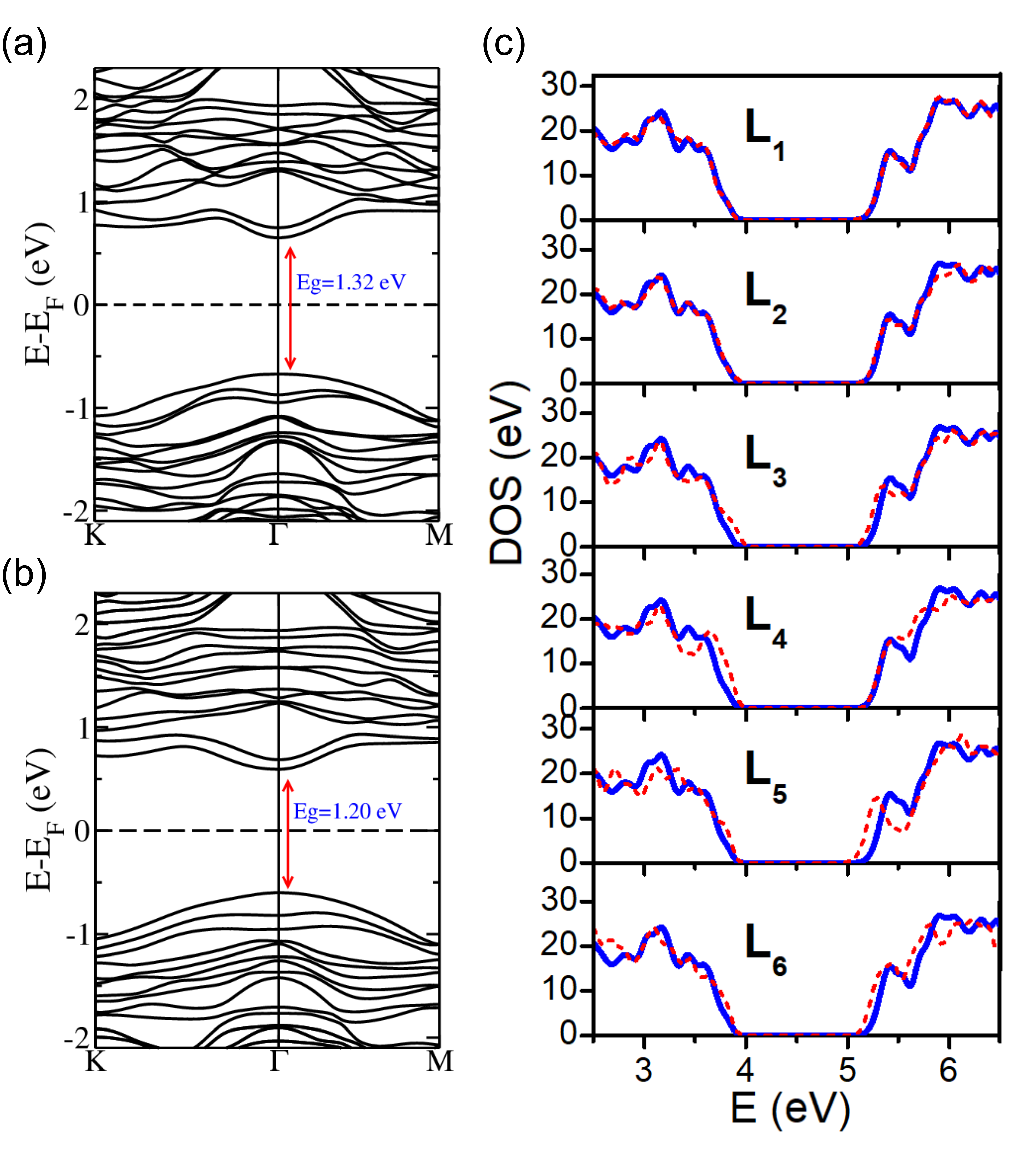}
		\caption{Electronic structures of bilayer ReS$_2$ in stacking configuration 0 obtained with (a) SOC = 0, (b) SOC $\neq$ 0. (c) Density of states of bilayer ReS$_2$ (blue solid line) in equilibrium and distorted structures (red dotted line) obtained by freezing various phonon modes. }
		
		\label{f6}
	\end{figure*}

	While hexagonal rings of ReS$_2$ structure are not ideal due to distortions of 1T-structure and formation of Re chains, we have used the nomenclature of hexagonal high symmetry points in our electronic structure. Electronic structure of bulk ReS$_2$  along high symmetry lines (see Fig. S10(a) of the SM) shows  an indirect band gap of 1.17 eV. Tongay \textit{et al.} \cite{Tongay} have reported a direct bandgap of 1.35 eV at $\Gamma$-point of bulk ReS$_2$ based on GGA calculations. To validate  our calculations, we carried out GGA calculations and confirm a direct bandgap of 1.51 eV at $\Gamma$-point and additionally we find a direct bandgap of 1.36 eV at A-point (Fig. S10(b)) which was previously not reported  by Tongay \textit{et al.} \cite{Tongay}. With inclusion of the spin-orbit coupling (SOC) in LDA-USPP calculations, we find a direct bandgap of 1.06 eV at A-point (Fig. S10(c)).

	Previous reports suggested three (two) stacking configurations for bilayer (bulk) ReS$ _{2} $ \cite{he2016coupling,doi:10.1002/adma.201908311}. Accordingly for our bilayer sample, we considered  stacking configurations (labeled stacking configuration 1, 2, 3 in Table-\ref{t2}) from Ref. \onlinecite{he2016coupling} and relaxed their
	structures. The energy difference between the various stacking configurations  has the same magnitude as reported by He \textit{et. al.} \cite{he2016coupling}. In addition, we have considered the fourth stacking configuration of bilayer ReS$_2$ with the two monolayers on top of each other, terming it as AA stacking (labeled stacking 0) similar to Ref. \onlinecite{doi:10.1002/adma.201908311}. Our results show that AA stacking is clearly the most stable configuration of all (Table-\ref{t2}). Secondly, calculated electronic 
	structures of the stacking 0 and stacking 3 configurations show a direct bandgap at $\Gamma$-point. For the stacking configuration 0 of bilayer ReS$_2$, we find a direct bandgap of 1.32 eV (without SOC) and  1.20 eV (with SOC) (Figs. \ref{f6}(a) and (b)). In bilayer ReS$_2$ with stacking  configuration 3, we find a marginally larger direct bandgap of 1.37 eV (without SOC) (Fig. S11(a)) and 1.23 eV (with SOC). We do not find a noticeable difference in electronic structure of these stacking configurations of bilayer ReS$_2$.
	
	\begin{table}[h!]
		\label{stack1}
		\centering
		\caption{ Energies of stacking configurations (n), (E$_n$-E$_0$, n=1,2,3), relative
			displacements (d) between the two layers of bilayer ReS$_2$ with respect to the bottom ReS$_2$, and direct band gaps at $\Gamma$-point
			of all the stacking configurations considered.}
		\begin{tabular}{c c c c}
			\hline
			\hline
			Stackings & Energy (meV) & d (\AA) & E$_g$ (eV) \\
			\hline
			
			0 & 0 & (-1.80, -0.87, 5.98) & 1.32 \\
			1 & 54 & (-0.008 2.17 6.05) & 1.35 \\
			2 & 80 & (-0.43 -0.11 6.19) & 1.37 \\
			3 & 49 & (0.06 -5.82 6.06) & 1.37 \\
			\hline
			\hline
		\end{tabular}
		\label{t2}
	\end{table}
	
	\begin{figure*}[ht!]	 
		\includegraphics[width=0.8\textwidth]{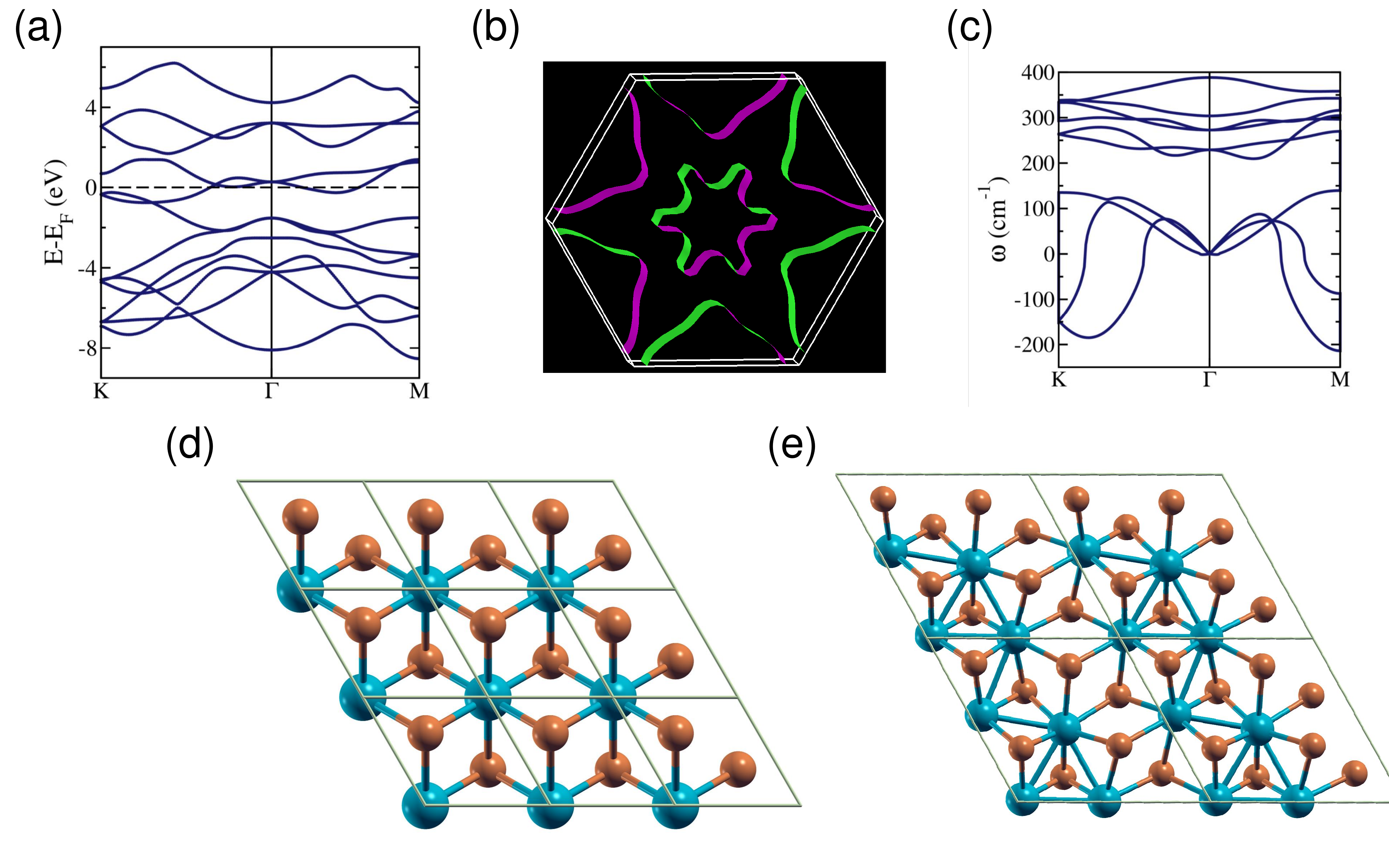}
		\caption{(a) Electronic structure of 1T-ReS$ _{2} $. (b) Fermi surface of 1T-ReS$ _{2} $ showing nesting. (c) Phonon dispersion of 1T-ReS$ _{2} $. (d) Top view of the structure of 1T-ReS$ _{2} $. (e) The intermediate structure formed after freezing the M-point phonon of 1T-ReS$ _{2} $.   }
		
		\label{f7}
	\end{figure*}
	
	For bilayer ReS$_2$, we simulated AA-stacking (stacking configuration 0) with an additional electronic charge equivalent to the experimental 
	concentration of electron doping. With electron doping, we find that  A$_{g}$-like modes undergo small changes in frequency whereas E$ _{g} $-like in-plane vibrational modes of relatively higher frequency exhibit a significant softening (Figs. {\ref{f2}}(c) and S2(d)). The trend in changes in calculated phonon frequencies with electron doping concentration qualitatively agrees with our experiments (Figs. \ref{f2}(a) and S2(c)), though calculated frequency shifts $\Delta\omega$ are higher than the experimental values. To understand these observed trends, we determined EPC of bilayer ReS$_2$ at different electron doping concentrations. We find
	that the EPC of E$ _{g}$-like modes (L$ _{3} $, L$_{4} $, L$ _{5} $ and L$ _{6} $)  increases with doping concentration whereas EPC of A$ _{g} $-like modes (L$ _{1} $ and L$ _{2} $) remain unaffected (Figs. \ref{f2}(d) and S2(d)). Moreover, the trend in calculated EPC is qualitatively similar to the  observed changes in phonon linewidth with doping (Figs. \ref{f2}(b) and S2(c)).

	From the density of electronic states of bilayer ReS$_2$ structure obtained after freezing various modes with amplitude of atomic displacements of 0.06 \AA, large shifts in the VBM and CBM positions are clearly evident due to the modes L$ _{3} $, L$ _{5} $ and L$ _{6} $, which couple strongly with the electrons (Fig. \ref{f6}(c)). However, shifts in the CBM and VBM are negligible for the out-of-plane modes (L$ _{1} $ and L$ _{2} $) confirming their weak coupling with electrons. We notice slight shifts in the CBM and VBM positions are associated with freezing of the mode  L$ _{4} $.

	1T$ ^{\prime} $ structure is a result of structural distortions of centrosymmetric, octahedrally coordinated 1T structure which is metallic (Figs. \ref{f7}(a) and (d)). Its nested Fermi-surface (Fig. \ref{f7}(b)), gives rise to unstable modes at M and $ \Gamma $ points (Fig. \ref{f7}(c)), that freeze in to generate the 1T$ ^{\prime} $ structure via an intermediate structure (Fig. \ref{f7}(e)). This essentially means that these unstable modes of the 1T structure couple strongly with its electronic states at the VBM and CBM. These are the modes involving in-plane displacements of Re and S atoms, which harden after transition to the 1T$ ^{\prime} $ structure. Indeed, they overlap strongly with the modes in 1T$ ^{\prime} $ structure that exhibit strong doping dependence (high EPC) observed in Raman experiments. Thus, the physical origin of this strong coupling is traced to the mechanism of nested Fermi surface making metallic 1T phase of ReS$ _{2} $ unstable, and stabilization of the observed semiconducting 1T$ ^{'} $ phase of ReS$ _{2} $ by the nesting phonon-related structural distortions involving in-plane displacement of the lattice.  This is the origin of strong coupling of these phonons with electronic states near the gap. As ReS$_2$ is vibrationally decoupled \cite{Tongay}, we have considered monolayer ReS$_2$ for the above analysis.

	Layer stacking sequence is known to influence shear modes of few layer ReS$_2$ \cite{qiao2016polytypism,lorchat2016splitting}, which can serve as a spectral signature of structural stacking. To predict how it affects phonon frequencies with electron doping, we obtained zone-center phonon spectrum for stacking 3 of bilayer ReS$_2$. We find that the trends of changes in frequencies and EPC remain the same (Figs. S11(b) and (c)). Thus, we conclude that the stacking sequences do not affect the doping dependence of high-frequency Raman modes of bilayer ReS$_2$, while they may influence shear modes.

	\section{Conclusions}
	In summary, we have carried out Raman measurement on bilayer ReS$ _{2} $ based field-effect transistor and developed an understanding of our results using DFT analysis. With electron doping, the in-plane Raman modes show phonon softening and linewidth broadening while no significant change is observed for the out-of-plane modes. The in-plane mode at $\sim$151 cm$ ^{-1} $ shows doping tunable Fano resonance and largest change in frequency and linewidth with electron doping. The Fano  parameter $ (1/q) $ reaches $\sim$ -0.17 at doping concentration of $\sim 3.7\times10^{13}$ cm$^{-2} $. Our DFT calculations show that the out-of-plane modes couple weakly with electrons while the in-plane modes couple strongly.  Atomic displacements of the in-plane modes of ReS$ _{2} $ account for the electronic and structural  transition from 1T to 1T$ ^{\prime} $, and hence these modes couple more strongly with electrons than the out-of-plane modes.  The stacking sequence of the bilayer ReS$_2$ does not alter their electronic properties significantly.Beside doping, external strain can also change the vibrational property at nanoscale \cite{doi:10.1021/nl402875m}. Therefore, we  hope our work will motivate further study to tune the Fano parameter with application of uniaxial strain along its two crystallographic axes, enabling ReS$ _{2} $ based devices to be used as a highly sensitive anisotropic strain sensor.  
	
	\section*{Conflicts of interest}
	There are no conflicts to declare.

	\section*{Acknowledgments}
	The authors would like to thank the Center for Nanoscience and Engineering (CeNSE), IISc for device fabrication facilities. U.V.W. acknowledges support from a J. C. Bose National Fellowship of DST, Government of India and AOARD project. AKS thanks Department of Science and Technology, Government of India for financial support under Nanomission and Year of Science Professorship.

	\bibliographystyle{apsrev4-2}
	\bibliography{ref}
	
\end{document}


\title{Supplemental Material\\Doping controlled Fano resonance in bilayer 1T$ ^{\prime} $-ReS$ _{2} $: Raman experiments and first-principles theoretical analysis}

\author{Subhadip Das$^1$, Suchitra Prasad$^2$, Biswanath Chakraborty$^{1,3}$, Bhakti Jariwala$^4$, Sai Shradha$ ^{4} $, D. V. S. Muthu$^1$, Arnab Bhattacharya$^4$, U. V. Waghmare$^2$ and A. K. Sood$^1$}

\email{asood@iisc.ac.in}
\affiliation{$^1$Department of Physics, Indian Institute of Science, Bangalore 560012, India\linebreak$^2$ Theoretical Sciences Unit, Jawaharlal Nehru Centre for Advanced Scientific Research, Bangalore 560064, India \linebreak $ ^{3} $ Present address: Department of Physics, Indian Institute of Technology Jammu, Jammu-181221,J\&K, India\linebreak $^4$Department of Condensed Matter Physics and Materials Science, Tata Institute of Fundamental Research, Mumbai 400005, India.}
\maketitle
\clearpage
\section{P\lowercase{olar plots of the out-of-plane} R\lowercase{aman modes}}
\begin{figure}[!ht]
	\includegraphics[width=0.8\linewidth]{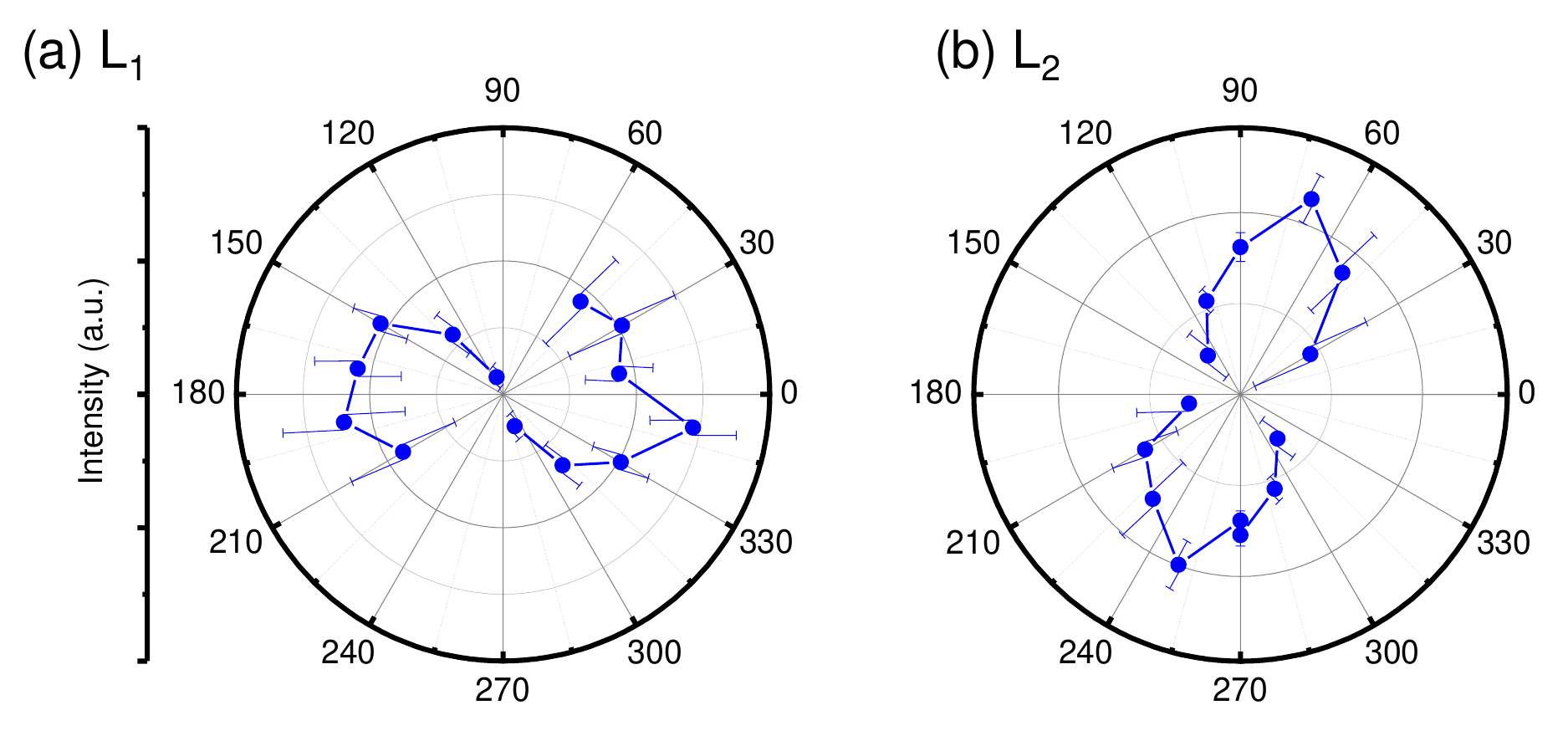}
	\caption{ 	(a) Raman intensity variation of the A$ _{g} $-like (a) L$ _{1} $ and (b) L$ _{2} $ modes  with laser polarization angle $\theta$ in unpolarized collection configuration. Intensities of the peaks have been normalized using angular dependence of Si peak at $\sim$520.5 cm$ ^{-1} $. }
	\label{s8}
\end{figure}
\newpage
\
\section{A\lowercase{dditional tranport data and} EPC \lowercase{analysis of 143 \lowercase{cm$ ^{-1} $} mode of bilayer} R\lowercase{e}S$ _{2}$ }
\begin{figure}[!ht]
	\includegraphics[width=0.84\textwidth]{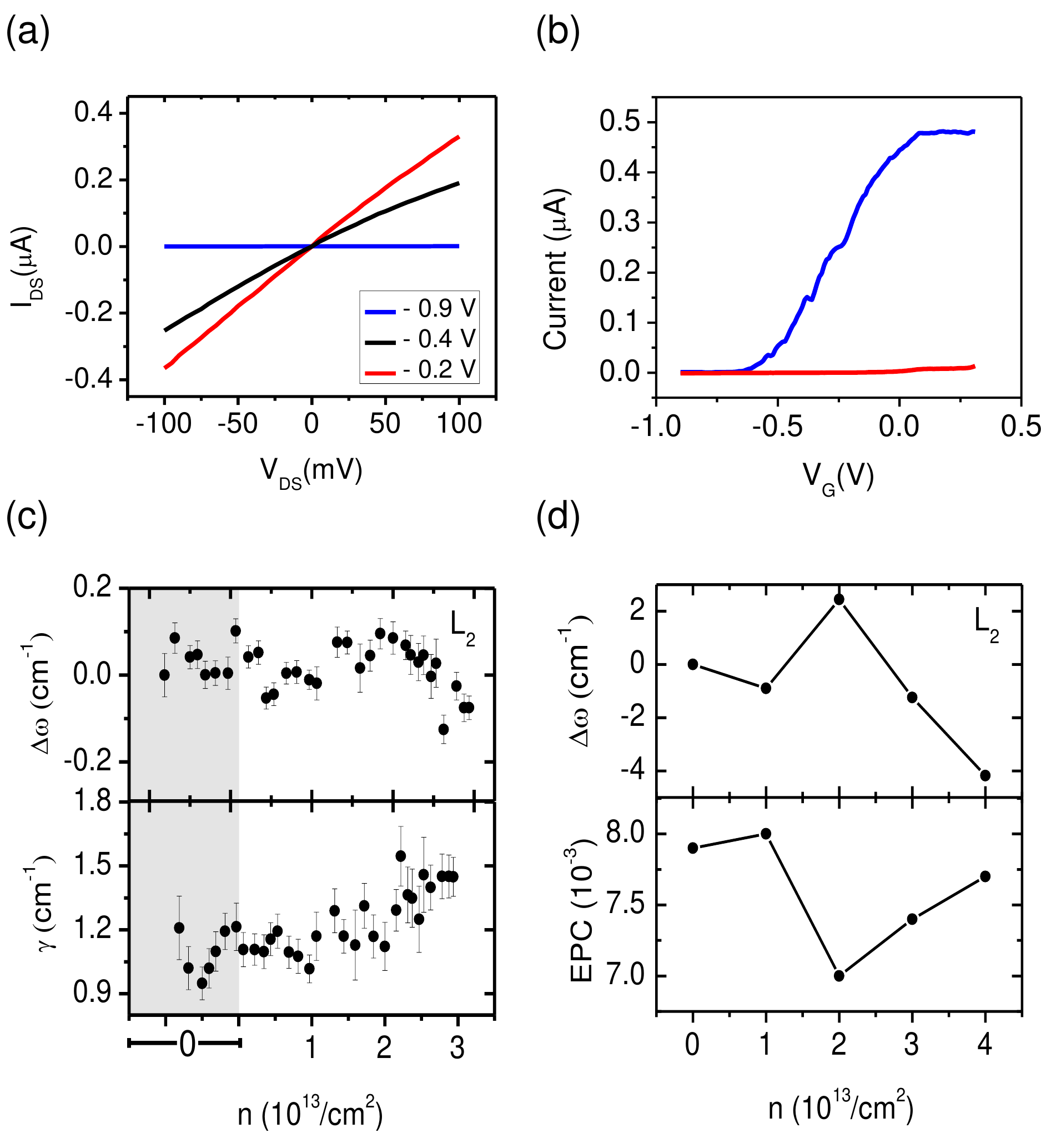}

	\caption{(a) Drain current (I$ _{DS} $) versus drain voltage (V$ _{DS} $) at three different gate voltages. (b) Negligible gate to source current (red line) compared to the drain to source (blue line) current at 0.1V of drain to source voltage, confirming the absence of Faradaic current in our device transport measurements from Figs. 1(d) and (e) of the main text.  (c) Experimental measurements of the frequency shift ($ \Delta\omega=\omega(n\neq0)-\omega(n=0)$) and linewidth ($ \gamma $) with electron doping (n) of the L$ _{2} $ mode at $\theta=90^{o}$. Gray region represents the transistor off-state (V$ _{G} \leq$V$ _{Th} $). (d) Calculated values of $\Delta\omega$ and EPC for the L$ _{2} $ mode with n.}
\end{figure}
\clearpage
\newpage
\section{B\lowercase{ehavior of the in-plane modes at $\theta=90^{o}$}}
\begin{figure}[!ht]
	\includegraphics[width=0.75\linewidth]{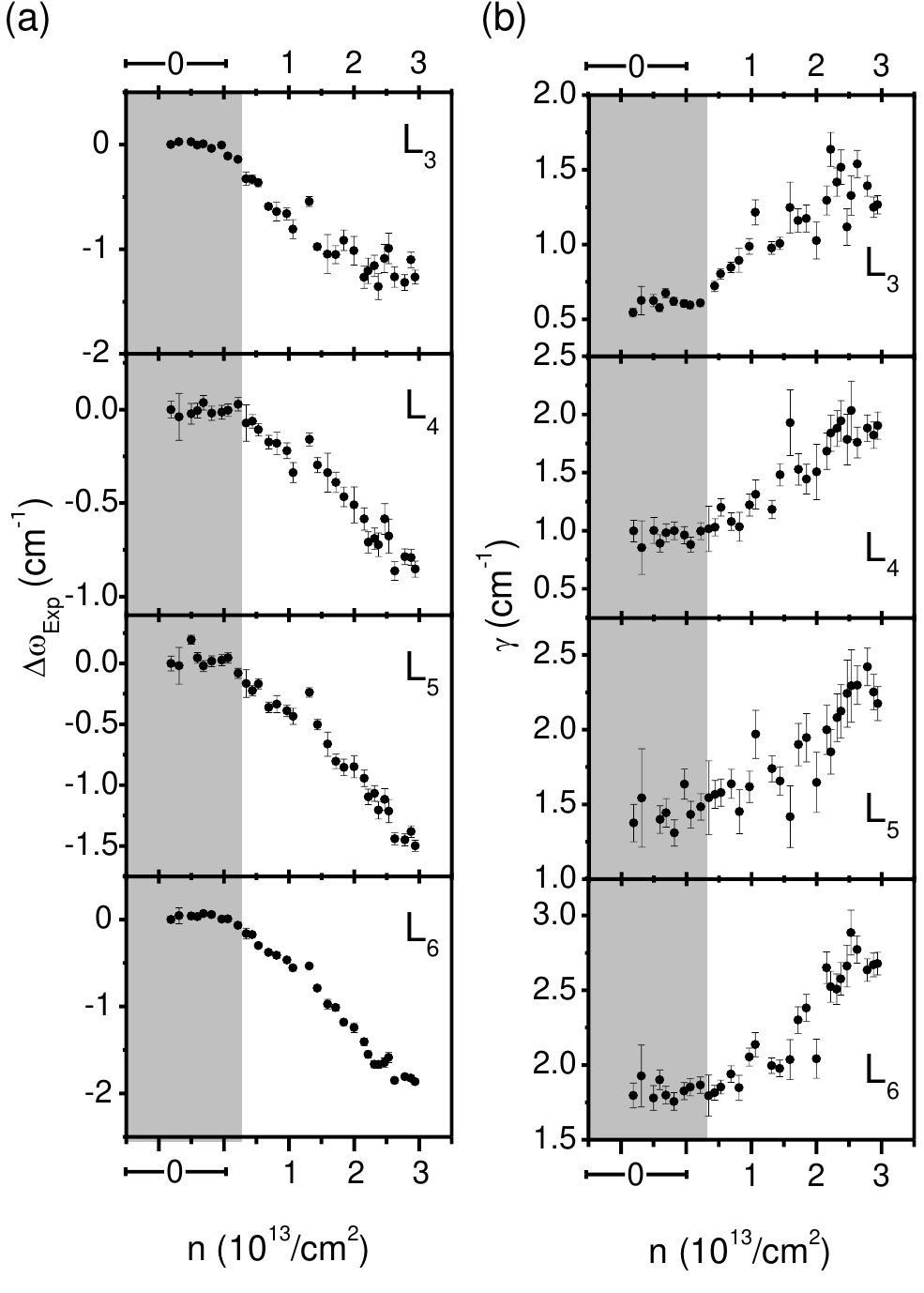}
	\caption{(a) $\Delta\omega$ and $\gamma$ versus n at $\theta=90^{o}$ for the E$ _{g} $-like modes.}
	\label{s11}
\end{figure}

\newpage
\section{EPC \lowercase{measurement of trilayer} R\lowercase{e}S\lowercase{e}$ _{2} $}

\begin{figure}[!htb]	
	\includegraphics[width=1\textwidth]{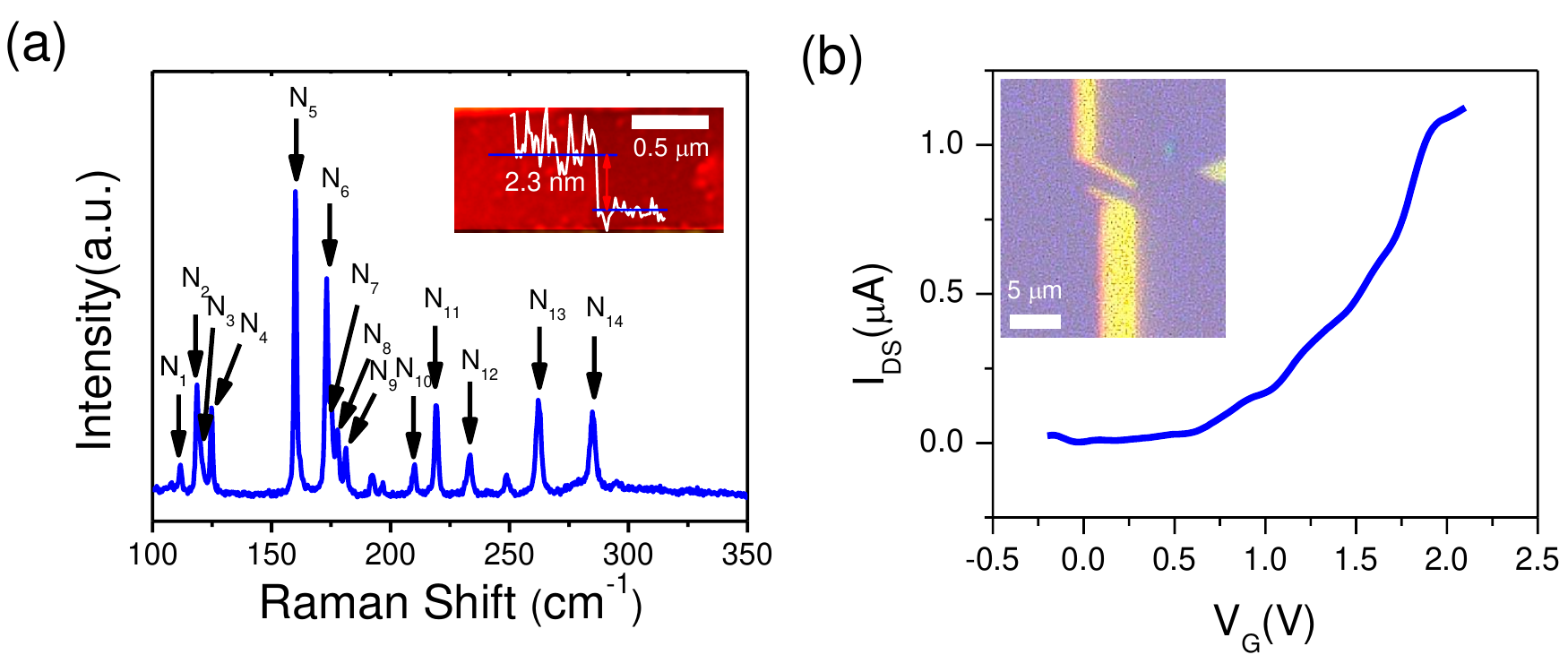}
		\caption{ (a) Raman spectrum of trilayer ReSe$ _{2} $. The modes are indicated as N$ _{1} $ to N$ _{14} $.  Inset image shows the AFM height profile.  (b) Drain current (I$ _{Ds} $) as a  function of the gate voltage (V$ _{G} $)  with drain voltage fixed at  0.4V. Inset shows the optical image of the device.   }
	\label{s2}
\end{figure}

\begin{figure}[!htb]
	
	\includegraphics[width=1\textwidth]{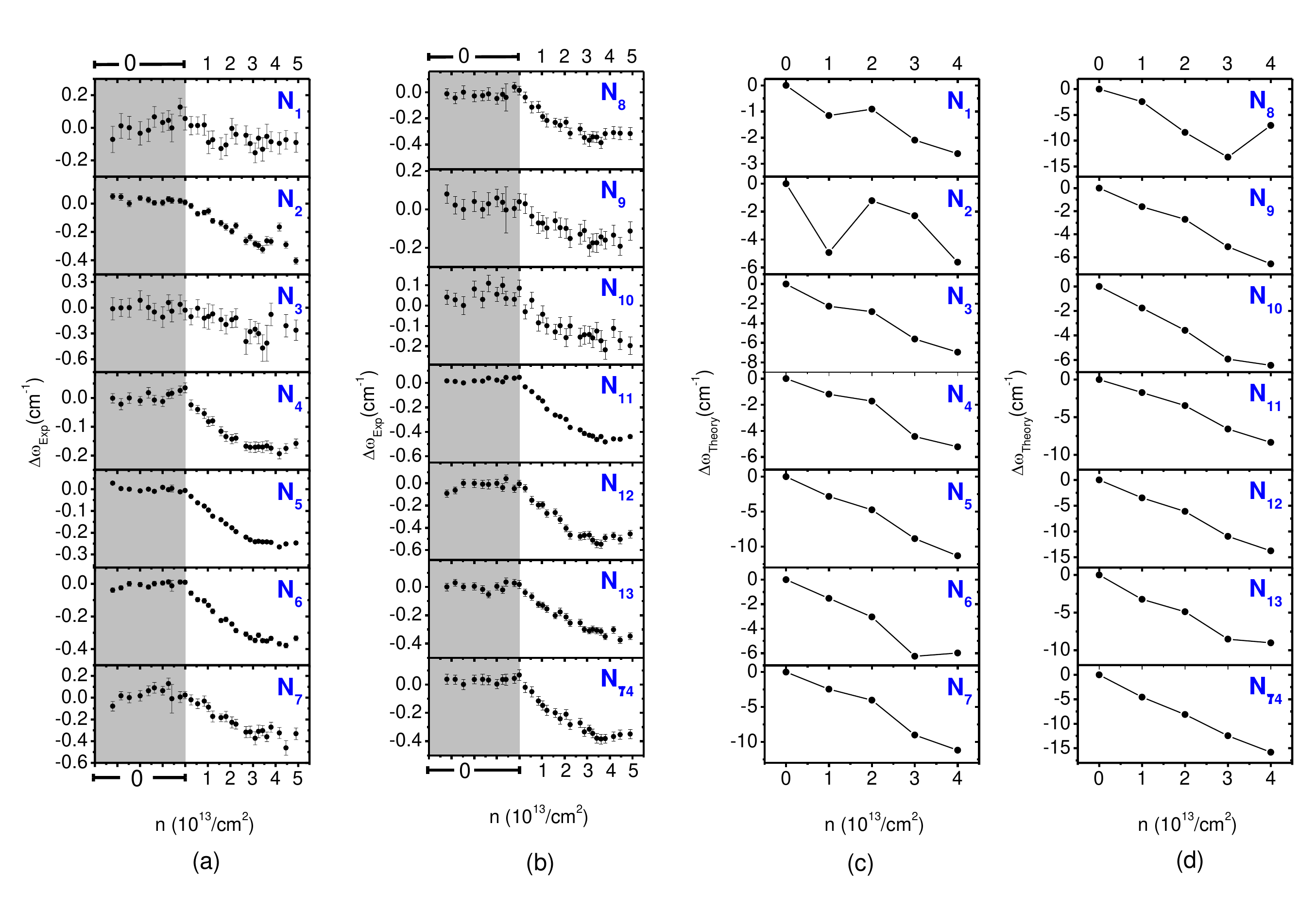}
	\caption{$\Delta\omega$ versus n from experimental measurements (a, b) and theoretical calculations (c, d) in trilayer ReSe$_2$. Gray region indicates undoped regime.} 
	\label{s3}
\end{figure}

\begin{figure}[!htb]
	\includegraphics[width=1\textwidth]{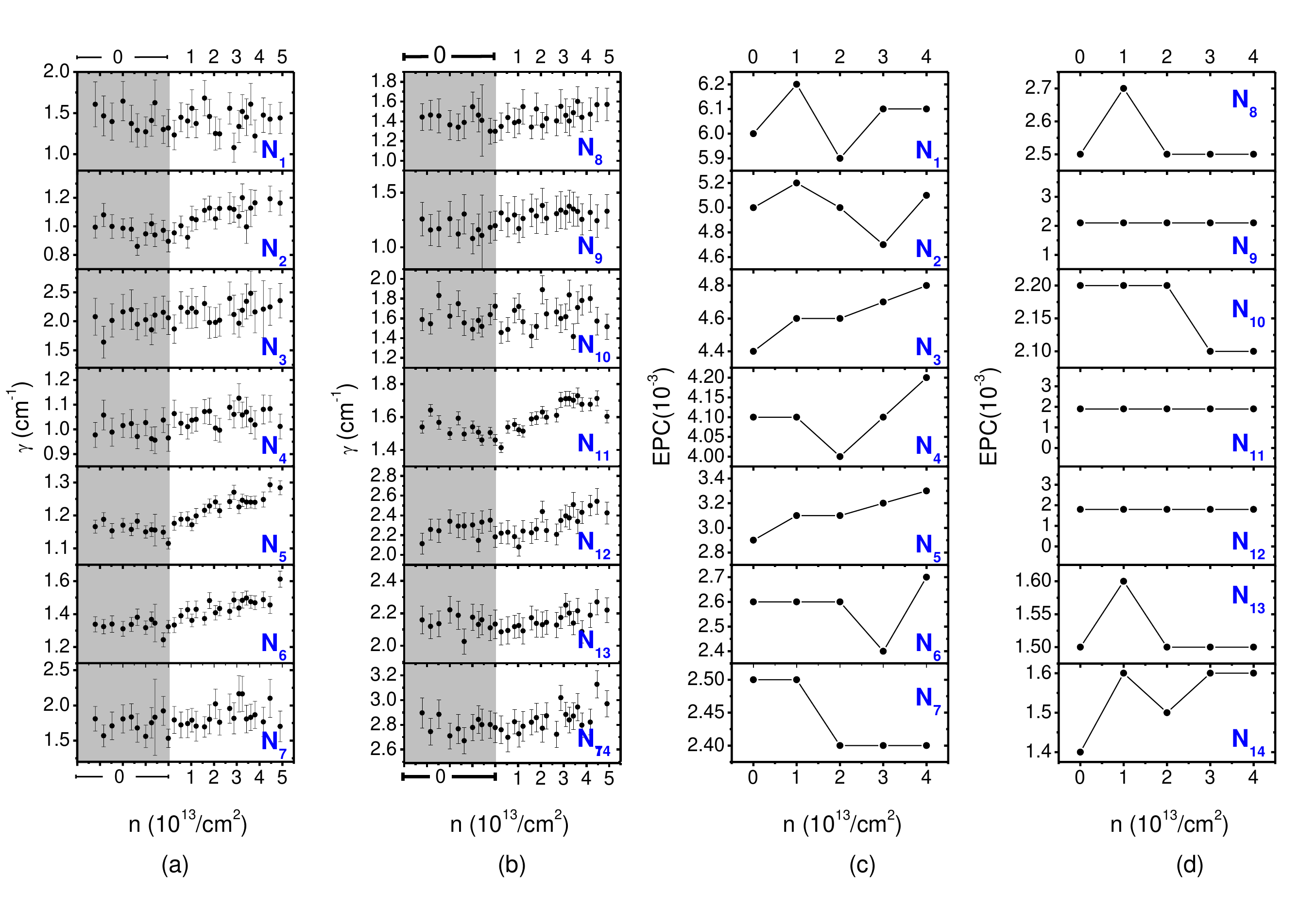}
	\caption{ Experimentally measured $\gamma$ (a, b)  and calculated EPC (c, d)  versus n. Gray region represents undoped regime. }
	\label{s4}
\end{figure}

Using similar methods for experimental measurements and theoretical calculations used for bilayer ReS$ _{2} $ (described in the main text), we have repeated our work on a trilayer ReSe$ _{2} $ transistor. ReSe$ _{2} $ is isostructural to ReS$ _{2} $ \cite{lamfers1996crystal,wildervanck1971dichalcogenides}, hence there are 18 Raman active mode with A$ _{g} $ symmetry \cite{wolverson2014raman}. Unlike ReS$ _{2} $, the Raman modes of ReSe$ _{2} $ has not been yet classified as in-plane or out-of-plane modes.  Atomic force microscopic measurement (inset of Fig. S\ref{s2}(a)) reveals $\sim$ 2.3 nm (3 layers) of sample thickness.  Raman spectrum for trilayer ReSe$ _{2} $ (Fig. S\ref{s2}(a)) shows 14 modes, which are labeled from N$ _{1} $ to N$ _{14} $. The drain current (I$ _{DS} $) with gate voltage (V$ _{G} $) measurement (Fig. \ref{s2}(b)) shows electron field-effect mobility of $\sim$ 0.41 cm$ ^{2} $/V.s and current on-off ratio of $\sim$ 10$ ^{2} $. Although  these parameters are consistent with previous reports \cite{jariwala2016synthesis,yang2014layer}, the  n-type semiconducting behavior can be attributed to unintentional doping during growth process \cite{jariwala2016synthesis}. The Raman modes of trilayer ReSe$ _{2} $ (Fig. S\ref{s2}(a)) at different gate voltages are fitted with a sum of Lorentzian functions to extract the phonon frequency ($\omega$) and linewidth ($\gamma$). All 14 modes show small changes in phonon frequency, with maximum phonon softening of $\sim$ 0.6 cm$ ^{-1} $ observed for N$ _{7} $, N$ _{11} $ and N$ _{12} $ modes at n $\sim$ 5$\times$10$ ^{13}$/cm$ ^{2}  $ (Figs. S\ref{s3}(a, b)). Similarly, the linewidth of all the modes (Figs. S\ref{s4}(a, b)) show little to no change with doping.

\begin{figure}[!htb]	
	\includegraphics[width=0.8\textwidth]{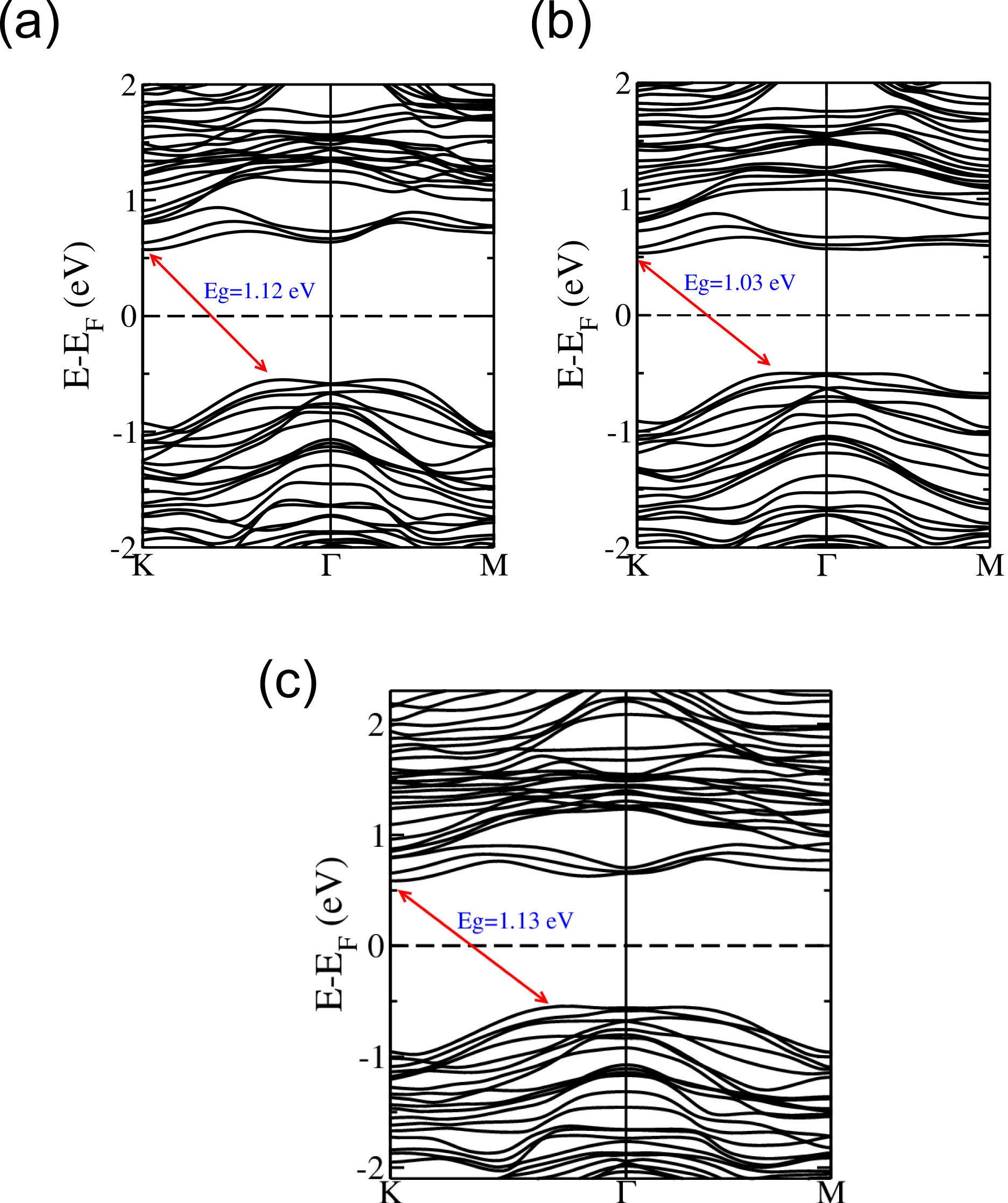}
	\caption{ Electronic structures of  trilayer ReSe$ _{2} $ in stacking configuration 0 obtained with (a) SOC =
		0, (b) SOC $\neq$ 0. (c) Electronic structure of trilayer ReSe$_2$ with stacking 2 obtained with SOC=0. }
	\label{s2b}
\end{figure}

 We find an indirect band gap of 0.93 eV and 1.02 eV of bulk
ReSe$_2$ based on the calculation with and without the inclusion of spin-orbit coupling (SOC). Our results are in 
good agreement with the earlier theoretical and experimental findings of band gaps of bulk ReSe$_2$ \cite{wolverson2014raman,zhao2015interlayer}. In trilayer ReSe$_2$, we have tried three different ABA stacking configurations. We displaced the 
middle layer of ReSe$_2$ by three different distances (d) and obtained the relative energy 
after \textit{z}-direction relaxation. We find that the AAA stacking is the most stable of all (Table-\ref{t3}).
For trilayer ReSe$_2$, we see 
an indirect band gap of 1.12 eV and VBM is a bit away from $\Gamma$-point (Fig. S\ref{s2b}(a)). 
After including the SOC in our calculations, we get an indirect band gap of 1.03 eV, 
with the VBM at $\Gamma$-point (Fig. S\ref{s2b}(b)). We determined the electronic structure 
of trilayer ReSe$_2$ in the ABA stacking configuration (stacking 2), which has lower
energy than the other ABA stacking configuration. We find an indirect band gap of 1.13 eV for 
ABA stacking of trilayer ReSe$_2$ (Fig. S\ref{s2b}(c)).
Also, the electronic structure does not 
change much with the stacking sequence. Since, AAA stacking (stacking 0) is the most stable 
of all, we used it in finding the effects of electron doping. We see negligible softening of 
phonon modes with electron doping (Figs. S\ref{s3}(c, d)). In addition, all the phonon modes show relatively weak coupling with electrons (Figs. S\ref{s4}(c, d)).

\begin{figure}[!htb]
	\includegraphics[width=0.8\linewidth]{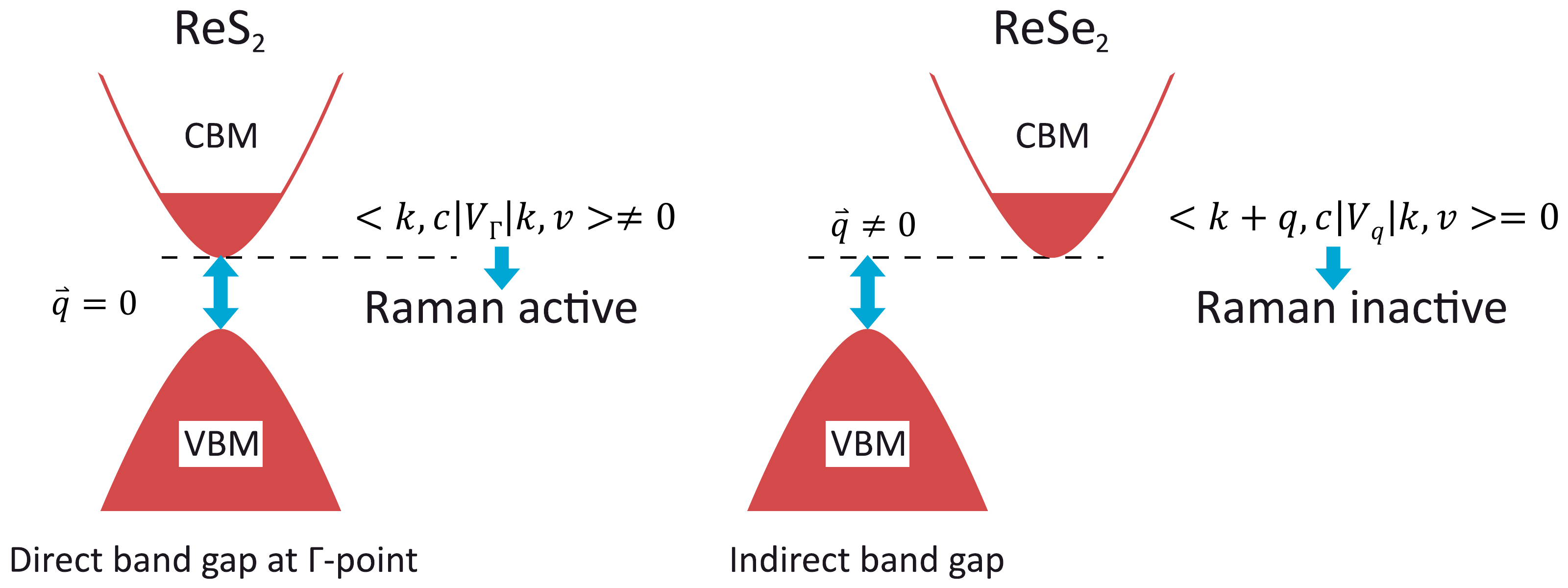}
	\caption{Schematic illustration of the coupling of electrons with phonons in ReS$_2$ and ReSe$ _{2} $.} 
	\label{s5}
\end{figure}

The direct band gaps of ReS$_2$ and ReSe$_2$ at $\Gamma$-point are 1.32 eV  and 1.22 eV respectively (Figs. 5(a) and S\ref{s2b}(a)). 
In contrast to ReS$_2$, ReSe$_2$ exhibits an indirect band gap of 1.12 eV (Fig. S\ref{s2b}(a)). The VBM of ReS$_2$ has the highest density of states at $\Gamma$-point while, it is slightly 
offset from $\Gamma$ in ReSe$_2$. Phonon frequencies are derived from the interatomic force 
constants which are linear response functions having a dominant contribution from the phonon mediated 
coupling between electronic states at CBM  and VBM.
Upon electron doping, the valley of CBM at
K-point of ReSe$_2$ gets populated, while that at $\Gamma$-point gets populated in ReS$_2$. 
As a result, the frontier states at $\Gamma$-point are masked from contributing to phonon 
frequencies in ReS$_2$ resulting in changes in dominant terms in the interatomic force constant. 
Hence upon electron doping, a significant change in phonon frequencies of ReS$_2$ are observed, 
while little changes are seen in ReSe$_2$ (Fig. S\ref{s5}).
\begin{table}[!htb]
	\centering
	\caption{Energies of stacking configurations (n), (E$_n$-E$_0$, N=1,2) and relative displacements (d)
		between the middle layer of ReSe$_2$ with respect to bottom layer of trilayer  ReSe$_2$.}
	\begin{tabular}{c c c}
		\hline
		\hline
		Stackings & Energy (eV) & d (\AA) \\
		\hline
		0 & 0 & (-0.67, -2.19, 6.35) \\
		1 & 0.34 & (-1.48, -0.77, 6.35) \\
		2 & 0.02 & (-2.30, 0.65, 6.35) \\
		\hline
		\hline
	\end{tabular}
	\label{t3}
\end{table}

\newpage
\clearpage
\newpage

\section{R\lowercase{aman spectra at two different spots}}
\begin{figure}[!ht]
	\includegraphics[width=0.7\linewidth]{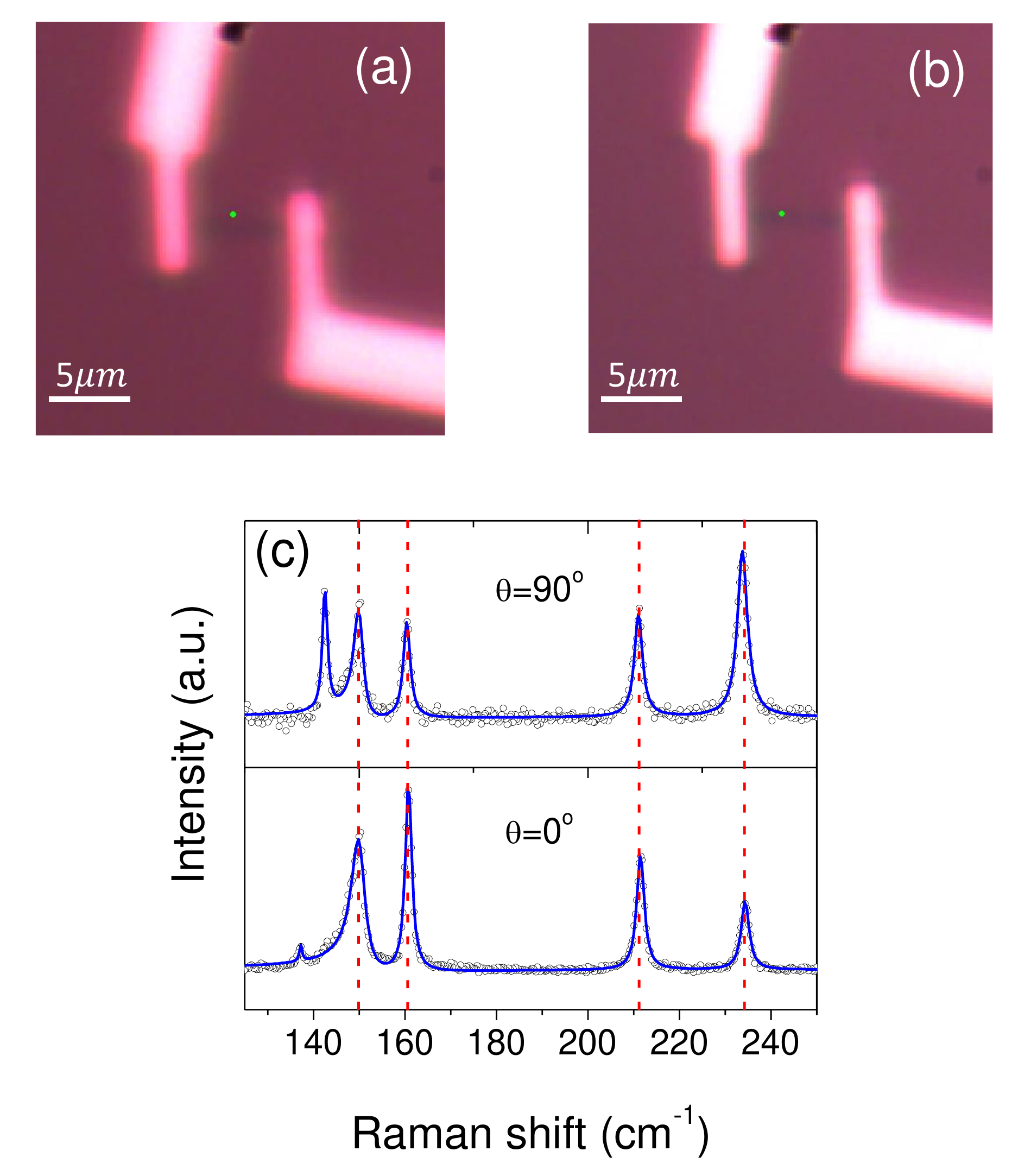}
	\caption{ Different incident laser positions shown by a green dot in the device optical images during (a) doping and (b) dedoping cycle. (c) Raman spectra with the same peak positions taken on these two spots at $ n\sim3\times10^{13}/cm^{2} $ during doping and dedoping cycles at $\theta=90^{o}$ and $ 0^{o} $, respectively. Black circles and blue lines are the experimental data points and their cumulative peak fits respectively.  As evident from the figure, only the mode at $\sim$ 153 cm$ ^{-1} $ (labeled L$ _{3} $ in Fig. 1(c)) shows asymmetric broadening in the low frequency side with the Fano parameter (1/q) of $\sim$ -0.15,  whereas other modes show symmetric Lorentzian lineshape. Red dashed lines are guide to the eye for Raman peak position for the in-plane modes.   }
	\label{s11}
\end{figure}
\clearpage
\section{B\lowercase{ulk band structure calculation of} R\lowercase{e}S$ _{2} $ }
\begin{figure}[!htb]
	\includegraphics[width=0.8\linewidth]{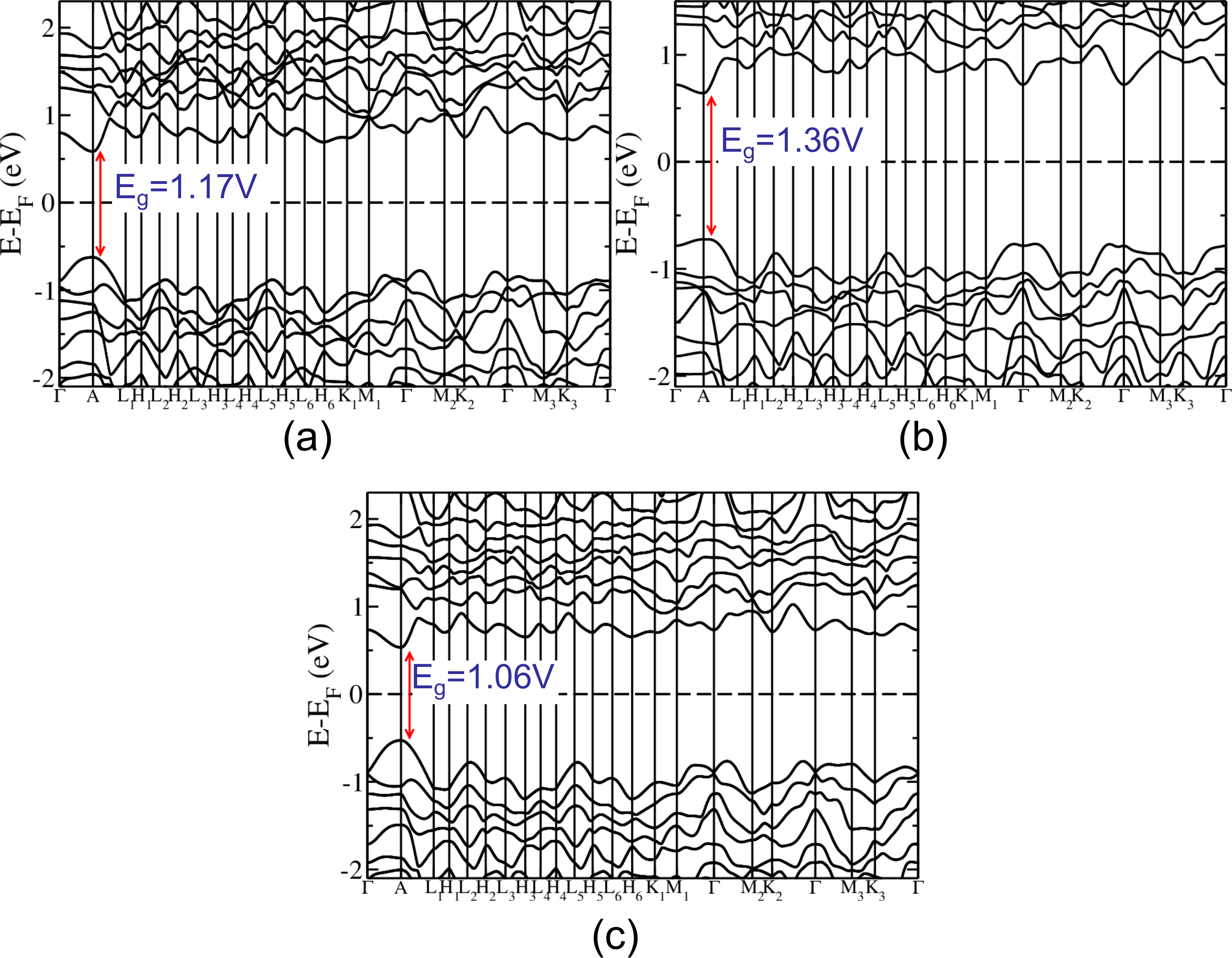}
	\caption{Electronic band structure of bulk ReS$_2$ obtained using (a) LDA-USPP,
		(b) GGA-USPP and (c) LDA-USPP with SOC inclusion.}
	\label{s6}
\end{figure}
\newpage
\pagebreak
\section{EPC \lowercase{calculation in stacking 3}}
\begin{figure}[!htb]
	\includegraphics[width=1\linewidth]{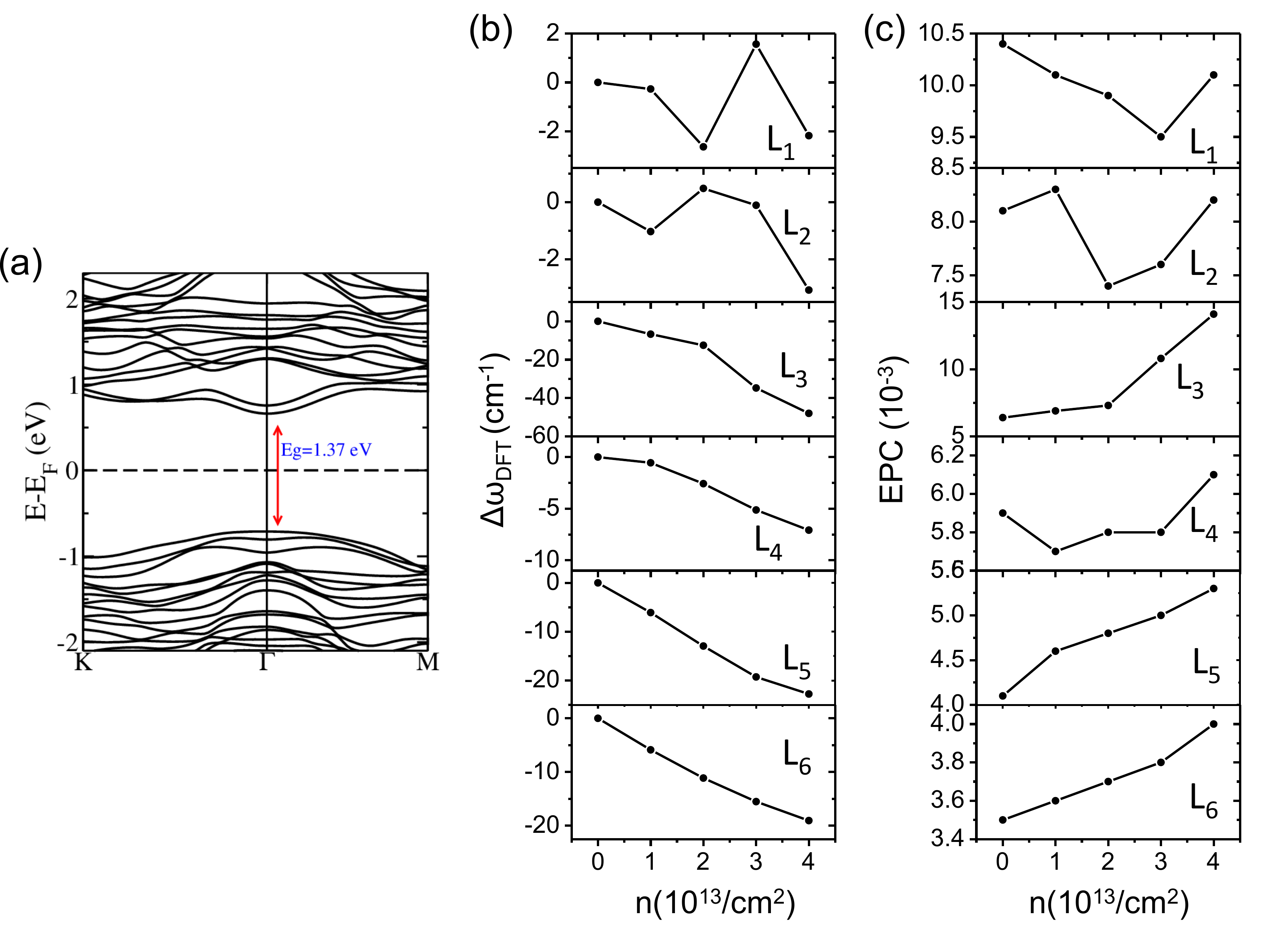}
	\caption{ 	(a) Electronic band structures of  bilayer ReS$_2$ with stacking 3. (b) Variation in phonon frequencies  and (c) electron-phonon coupling with electron doping concentration in bilayer ReS$_2$ (stacking 3), obtained from first-principles DFT calculations.}
	\label{s7}
\end{figure}

\bibliography{ref}